\documentclass[conference]{IEEEtran}
\pdfoutput=1

\usepackage{dsfont}
\usepackage{float}
\usepackage{graphicx}
\usepackage{multirow}
\usepackage{mwe}
\usepackage{array}
\usepackage{amssymb}
\usepackage{soul}
\usepackage{color}
\usepackage{url}
\newcolumntype{P}[1]{>{\centering\arraybackslash}p{#1}}

\begin{document}
\title{SoundCLR: Contrastive Learning of Representations For Improved Environmental Sound Classification}

\author{\IEEEauthorblockN{Alireza Nasiri}
\IEEEauthorblockA{Department of Computer Science and Engineering \\
University of South Carolina\\
Columbia, SC 29201, USA \\
Email: anasiri@email.sc.edu}
\and
\IEEEauthorblockN{Jianjun Hu}
\IEEEauthorblockA{Department of Computer Science and Engineering \\
University of South Carolina\\
Columbia, SC 29201, USA \\
Email: jianjunh@cse.sc.edu}}

\maketitle

\begin{abstract}
Environmental Sound Classification (ESC) is a challenging field of research in non-speech audio processing. Most of current research in ESC focuses on designing deep models with special architectures tailored for specific audio datasets, which usually cannot exploit the intrinsic patterns in the data. However recent studies have surprisingly shown that transfer learning from models trained on ImageNet is a very effective technique in ESC. Herein, we propose SoundCLR, a supervised contrastive learning method for effective environment sound classification with state-of-the-art performance, which works by learning representations that disentangle the samples of each class from those of other classes. Our deep network models are trained by combining a contrastive loss that contributes to a better probability output by the classification layer with a cross-entropy loss on the output of the classifier layer to map the samples to their respective 1-hot encoded labels. Due to the comparatively small sizes of the available environmental sound datasets, we propose and exploit a transfer learning and strong data augmentation pipeline and apply the augmentations on both the sound signals and their log-mel spectrograms before inputting them to the model. Our experiments show that our masking based augmentation technique on the log-mel spectrograms can significantly improve the recognition performance. Our extensive benchmark experiments show that our hybrid deep network models trained with combined contrastive and cross-entropy loss achieved the state-of-the-art performance on three benchmark datasets ESC-10, ESC-50, and US8K with validation accuracies of 99.75\%, 93.4\%, and 86.49\% respectively. The ensemble version of our models also outperforms other top ensemble methods. The code is available at https://github.com/alireza-nasiri/SoundCLR

\end{abstract}


%
\IEEEpeerreviewmaketitle

\section{Introduction}
Sound classification has a wide variety of applications in robot navigation \cite{li2007robot}, surveillance systems \cite{radhakrishnan2005audio}, alert systems \cite{intani2013crime}, damage monitoring of materials \cite{nasiri2019online}, wildlife monitoring \cite{chu2009environmental}, \cite{cakir2017convolutional}, and designing autonomous cars \cite{dobre2015low}, \cite{foggia2015reliable}. Environmental sounds have much higher variety than speech and this diversity and their noise-like characteristics makes ESC much more challenging than speech recognition. However, in recent years, sound recognition has seen great progress, which is partially due to the availability of large-scaled labeled datasets such as Environmental Sound Classification (ESC-50, ESC-10) \cite{piczak2015esc}, and Urbansound8k(US8K) \cite{salamon2014dataset}. The other reason is due to the shift from the traditional machine learning methods to deep learning models in the sound analysis tasks \cite{hershey2017cnn}, \cite{nasiri2019audiomask}. Signal processing and machine learning methods such as matrix factorization \cite{mesaros2015sound}, \cite{bisot2016acoustic}, dictionary learning \cite{salamon2015unsupervised}, Support Vector Machines (SVM) \cite{chu2009environmental}, \cite{uzkent2012non}, Gaussian Mixture Model (GMM) \cite{dhanalakshmi2011classification}, \cite{mesaros2017detection}, and K-Nearest Neighbor (KNN) \cite{piczak2015esc}, \cite{da2019evaluation}, are widely adopted in ESC. However, currently deep learning models mostly consisting of convolutional layers, are achieving the highest accuracies in ESC benchmark studies \cite{guzhov2020esresnet}, \cite{tokozume2018learning}, \cite{abrol2020learning}.

More recently, there has been studies that utilize transfer learning to improve ESC with significant improvements \cite{guzhov2020esresnet}, \cite{palanisamy2020rethinking}. These studies employ the deep fully convolutional models, which are, surprisingly, pre-trained on ImageNet, and fine-tune them using the spectrogram representation of the audio samples. Since spectrograms of the audio events show some kind of local correlation similar to the images, models pre-trained on the large image datasets, perform well in the audio classification too.

In most supervised learning approaches for audio classification, the training is guided by the cross-entropy loss \cite{guzhov2020esresnet}, \cite{palanisamy2020rethinking}, \cite{piczak2015environmental}, \cite{zhang2019learning}. The cross-entropy loss measures the difference between the output probability distribution and 1-hot encoded label and the model tries to reduce this loss by mapping the samples of the same class to their respective label vector.

Since a cross-entropy loss does not explicitly push away the samples of different classes from each other, it can suffer from poor margins between the representations of the samples from different classes \cite{elsayed2018large}. This problem can negatively affect the generalization power of the model. We argue that we can strengthen the training signal by explicitly disentangling the samples of different classes from each other in the representation space. This can be done by introducing another term into the loss function which tries to pull the samples of the same class together while pushing them away from the samples of other classes in the representation space. The supervised contrastive loss \cite{khosla2020supervised} can be a suitable candidate for this.

The contrastive loss function which recently made significant progress in self-supervised learning \cite{oord2018representation}, \cite{he2020momentum}, \cite{hjelm2018learning}, is based on defining positive and negative pairs by which it aims to pull together the samples in the positive pair (different augmented views of the same sample), while pushing away the samples in the negative pair (augmented views of different samples in the dataset). In self-supervised learning, the goal is to create a strong representation for the samples without using the ground-truth labels. By forcing the model to distinguish the different views of the same sample from the views of other samples, the model learns to focus on the most discriminative features of the samples.

Khosla et al. \cite{khosla2020supervised} has introduced a modified version of the contrastive loss for image classification, where the actual labels are employed to generate the positive and negative pairs. In the so-called supervised contrastive loss function, the samples of the same class are being pulled together in the representation space, while being pushed away from the representations of the samples from the other classes. In \cite{khosla2020supervised}, they propose to train a model using supervised contrastive loss as a feature extractor in the first stage, and then freeze the weights of the model in the second stage and train a shallow classifier on top of it using cross-entropy loss. This model achieves higher accuracy than the same neural network model trained using the cross-entropy loss for image classification. Inspired by this study, here we aim to examine the effect of the supervised contrastive loss for ESC. In this paper, We propose a new framework for ESC by introducing the supervised contrastive loss as the complementary to the cross-entropy loss for ESC. By calculating the training signals using both loss functions simultaneously, we allow the network parameters of the feature extraction model to be adjusted by a stronger signal with a single-stage training process.

In this paper, we propose SoundCLR, a new framework for ESC which is based on strong data augmentation, supervised contrastive loss on the representation space along with the cross-entropy loss on the final output of the model. We show that by adding the cross-entropy loss as a factor to the supervised contrastive loss, we can further separate audio event classes from each other and reach higher accuracy than the exact same models trained by either cross-entropy or supervised contrastive loss alone. Due to the small sizes of environmental sound datasets \cite{piczak2015esc}, \cite{salamon2014dataset}, applying data augmentation techniques is crucial in achieving high accuracy in ESC. Here in addition to applying the common strong data augmentation methods, we introduce a simple but effective masking based data augmentation technique for ESC. Extensive experiments over three common environmental sound benchmark datasets show that our new framework achieves the state-of-the-art performance when compared to existing independent and ensemble models. Also, we show that increasing the number of channels by just triplicating the input mel-spectrograms can improve the accuracy of the classifiers trained on ImageNet.

Our major contributions in this paper are as follows:  
\begin{enumerate}
\item We propose a new framework for environmental sound classification based on a hybrid loss function consisting of a supervised contrastive loss and a cross-entropy loss. The supervised contrastive loss is applied on the representation space to disentangle the samples of the classes from each other, and the cross-entropy loss is applied on the output of the classifier to map the representation vectors of the samples to their respective ground-truth labels.
\item We build a strong data augmentation pipeline, modifying the input data in both the wave signals and the mel-spectrograms. We show that triplicating input mel-spectrograms when using models pre-trained on ImageNet and random-masking of the mel-spectrograms can significantly improve the results.
\item  We show that the deep neural network models trained with our proposed hybrid loss outperform the models with the same architecture, but trained by using only either the cross-entropy loss or the supervised contrastive loss. Also, we show that our model achieves state-of-the-art results on ESC-50, ESC-10 and US8K datasets.
\end{enumerate}

The rest of the paper is organized as: In section \ref{related_work}, we briefly review the current approaches for sound classification, and then discuss the cross-entropy and contrastive loss in more details. Section \ref{proposed_method} goes over the different components of our proposed method. Experimental setup and their results are shown in section \ref{experiments}. Finally we conclude the paper in section \ref{conclusion}.

\section{Related Work}
\label{related_work}
\subsection{Environmental Sound Classification}
Environmental sound classification (ESC) deals with identifying some of the everyday audio events with varying lengths in a given audio signal. Even though a wide range of machine learning methods are proposed for sound classification \cite{piczak2015esc}, \cite{dhanalakshmi2011classification}, \cite{mesaros2017detection}, the deep learning methods have proven to achieve the best results in this field in recent years.

One of the earliest deep learning models for ESC was introduced by Piczak \cite{piczak2015environmental}. He used a 2-D structure, derived from the log-mel features of the audio signal as input to a model with two convolutional layers and two fully connected layers. This small model achieved 64.5\% accuracy on ESC-50, and 81.0\% on ESC-10, which is an increase of 20.6\% for ESC-50 and 7.8\% for ESC-10 compared to other traditional machine learning approaches such as random forest\cite{piczak2015esc}. In the following studies, the researchers developed deeper convolutional models which achieve even higher accuracies in ESC \cite{zhu2018learning}, \cite{tokozume2017learning}. In \cite{tokozume2017learning}, a model made of a combination of 1-D convolutional layers and fully connected layers extracts the features from the raw waveforms and achieves 71.0\% accuracy on ESC-50. Zhu et. al \cite{zhu2018learning} proposed a model with six convolutional layers to extract features from both the raw waves and the spectrograms. They gain accuracy values of 93.75\% and 79.1\% on ESC-10 and ESC-50, respectively.

Training deep neural networks with millions of parameters needs large amount of data. Current ESC benchmark datasets are considered to be comparatively small in the deep learning paradigm. Applying data augmentation techniques partially addresses this challenge \cite{salamon2017deep}, \cite{zhang2018deep}, \cite{tokozume2018learning}. Salamon et. al extensively studied the effects of data augmentation techniques like time-stretching, pitch-shifting  and adding background noise on improving the performance of their proposed CNN model \cite{salamon2017deep}. 
Zhang et. al proposed using mixup on audio signals to train a model of stacked convolutional and pooling layers \cite{zhang2018deep}. Similar to mixup, between-class learning proposed in \cite{tokozume2018learning}, mixes signals of the samples from different classes according to a random ratio, and a deep model of convolutional layers is trained to output the mixing ratio. By outputting the mixing ratio, the model learns the most important features in the sound signal. They achieve accuracy of 84.9\%, 91.4\%, and 78.3\% over three datasets ESC-50, ESC-10, and US8K, respectively.  

It is clear that deeper convolutional models and data augmentation  can both improve the ESC classification performance. However the sizes of the common environmental datasets put a constraint on training models with around fifteen convolutional layers. To train deeper models that can potentially achieve better results, transfer learning with models pre-trained on ImageNet has proven to be successful for ESC\cite{guzhov2020esresnet}, \cite{palanisamy2020rethinking}, \cite{palanisamy2020rethinking}. Since spectrograms, which are commonly used to train audio classifiers, show image-like characteristics such as close correspondence between the local points, using the pre-trained models on ImageNet contributes to a better feature-extraction and in turn to a higher classification accuracy. More recently, ESResNet-Attention \cite{guzhov2020esresnet} uses a ResNet-50 model pre-trained on ImageNet with parallel attention blocks for both frequency and time domains. It achieves accuracies of 97\%, 91.5\%, and  85.4\% on ESC-10, ESC-50 and US8K. Also, Palanisamy et. al \cite{palanisamy2020rethinking}, studied the performance of different well-known pre-trained models on ImageNet in audio classification. They report the validation accuracies of 90.65\% and 84.76\% on ESC-50 and US8K data sets when they use a single ResNet model. However, their best results comes from an ensemble of 5 DenseNet models where they achieve 92.89\% and 87.42\% accuracies on ESC-50 and US8K. They set a new state-of-the-art performance on these datasets.

\subsection{Cross-entropy and Contrastive Loss Functions}

All of the classifiers in the previous studies in ESC, use the cross-entropy loss to train their deep neural network models \cite{guzhov2020esresnet}, \cite{tokozume2018learning}, \cite{palanisamy2020rethinking}, \cite{zhang2018deep}. Cross-entropy loss was introduced to train neural networks with probability outputs \cite{rumelhart1986learning}, \cite{baum1987supervised}, and now is one of the most commonly used loss functions in classification tasks. This loss function measures the entropy between the actual probability distribution of the samples and the output probability distribution of the model. It can be calculated as:
\begin{equation}
    H(p, q) = -\sum_{x}p(x) log(q(x))
\end{equation}
where $p$ is the true probability distribution and $q$ is the calculated one. By training the model to minimize the cross-entropy loss, the samples in the same class are mapped to the near-by points in the embedding space and the intra-class distances are minimized. 

One of the downsides of the cross-entropy loss is the poor margins between the samples of different classes \cite{elsayed2018large}, which reduces the generality of the model trained by this loss function. The poor margins, or low inter-class distances, are related to the fact that there is no term in current loss function to push samples of different classes away from each other in the representation space.

There are a few loss functions proposed to improve the discrimination power of classifiers. Soft nearest neighbor loss \cite{salakhutdinov2007learning}, \cite{frosst2019analyzing}, and triplet loss \cite{hoffer2015deep} leverage the euklidean distances among the representation vectors of the samples to separate different classes from each other in the representation space. In both of them, minimizing the loss will force the model to output values which would minimize the intra-class to inter-class distance ratios. The main difference is that in the soft nearest neighbor loss, these distances are calculated among all of the samples present in the batch, where the triplet loss only considers 3 samples at a time: anchor, positive, and negative, where anchor and positive have the same label and the negative sample has a different label.

Another related work is contrastive learning \cite{hadsell2006dimensionality}, which has recently seen a lot of progress in the self-supervised learning field \cite{oord2018representation}, \cite{he2020momentum}, \cite{hjelm2018learning}, \cite{li2020prototypical}. Their main idea is the contrastive loss, which is based on building groups of samples as positives and negatives. It aims to decrease the distances among the representations of the samples in the positive group while increasing the distance among the representations of the samples in the negative group. It trains the model by pulling the representations of the samples from the positive pair together and pushing the representations of the samples in negative pairs, away from each other. In self-supervised learning \cite{he2020momentum},\cite{li2020prototypical}, \cite{chen2020simple}, this loss is applied to pull together embeddings of two augmented views of a single data sample at a time, while pushing them away from the augmented views of the other samples. In the supervised version of the contrastive loss proposed by \cite{khosla2020supervised}, instead of augmented views of the same sample, the positive pairs are built using samples with the same labels. This is very similar to what the triplet loss calculates. The major distinction is that the triplet loss only uses one negative and one positive pair while in supervised contrastive learning, we can have more than two samples in the positive or the negative group, which improves the training gradients by building a stronger contrast among the multiple samples and then may achieve higher accuracy. The proposed supervised contrastive learning \cite{khosla2020supervised} consists of two stages. In the first stage, the supervised contrastive loss is applied to disentangle the samples of the different classes from each other and in the second stage, the model's parameters are frozen and a classifier is trained on top of the model using the cross-entropy loss. Here, the supervised contrastive loss is introduced as an alternative to the cross-entropy loss. However, it seems that these two loss functions can be applied together as they are complementary to each other, where the representation space is simultaneously divided between the samples of the different classes using the contrastive loss, and the representation vectors are mapped to the ground-truth labels using the cross-entropy loss. In this paper, we investigate whether the contrastive loss can train better models  for environmental audio classification and propose a simple framework, in which both contrastive and cross-entropy loss functions are applied together to train a classifier, which achieves higher accuracy than the models trained with either one of these loss functions.

\section{Proposed Method}
\label{proposed_method}

\begin{figure}[h]
  \centering
  
  \includegraphics[width=0.45\textwidth]{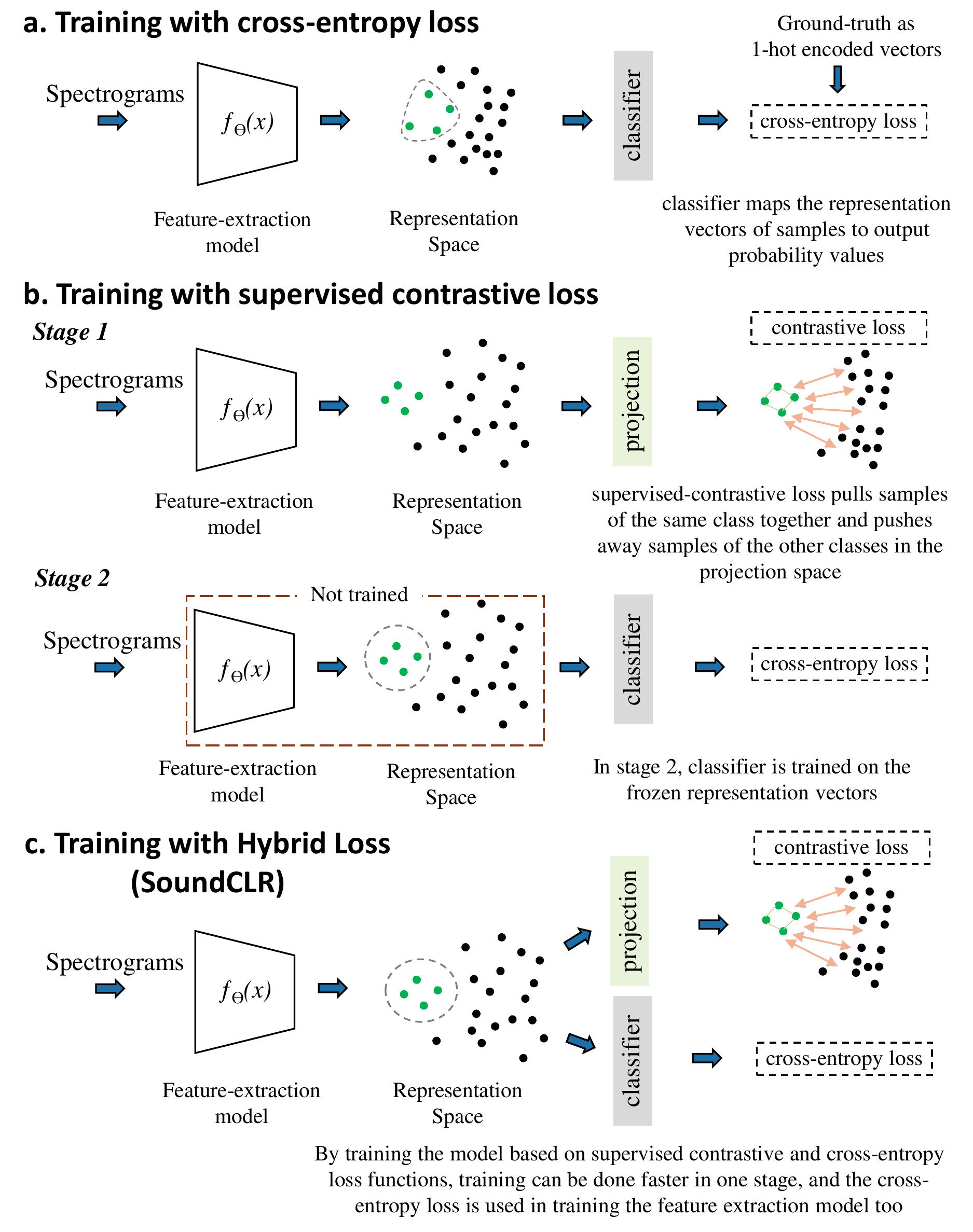}
  \caption{Overview of our three training schemes: a. model is trained with cross-entropy loss which measures the difference between the output probabilities and the 1-hot encoded ground-truth labels, b. training a classifier with supervised contrastive loss within two stages. In the first stage, the supervised contrastive loss disentangles the samples of different classes from each other in the representation space, and in the second stage, the feature-extraction model is frozen and a classifier layer is trained on top of the representation vectors. Here, the projection layer is only used for dimension reduction in stage 1, since applying the contrastive loss directly on the high-dimensional representation space is not as effective in separating the samples, c. training scheme using our proposed hybrid loss (SoundCLR), which utilizes both cross-entropy and supervised contrastive loss functions to disentangle the samples and map them to their corresponding labels, simultaneously. }
  \label{fig:overview}
\end{figure}

Figure \ref{fig:overview} shows an overview of our three implemented models, where the first model is trained using the cross-entropy loss, the second model is trained using the supervised contrastive loss and the cross-entropy loss in two stages, and finally our proposed hybrid models are trained with both cross-entropy and supervised contrastive loss simultaneously within one stage. 

In this remainder of this section, we present the different modules of our method in details. We start by specifying our data augmentation techniques in \ref{augmentation}. Then we discuss the characteristics of the three methods that we use to train the classifiers for environmental sounds. In \ref{baseline}, we go over the details of our baseline model which is trained using the cross-entropy loss. In \ref{hybrid}, we explain the components of our audio classifier trained with the contrastive loss in two steps, and we present our hybrid loss, which uses the contrastive loss to disentangle the representation space while simultaneously using the cross-entropy loss to map the representations to their respective ground-truth labels.

\subsection{Data Augmentation}
\label{augmentation}
In the small datasets like ESC-50 and ESC-10, data augmentation plays a big role in improving the performance of deep neuralal network models. Data augmentation is a collection of deformations applied to the signal that creates a slightly different signal while preserving its ground-truth label. This technique exposes the model to a wider variety of samples of a class and as a result, improves the generalization power of the model.

\begin{figure*}[h]
  \centering
  \includegraphics[width=\textwidth]{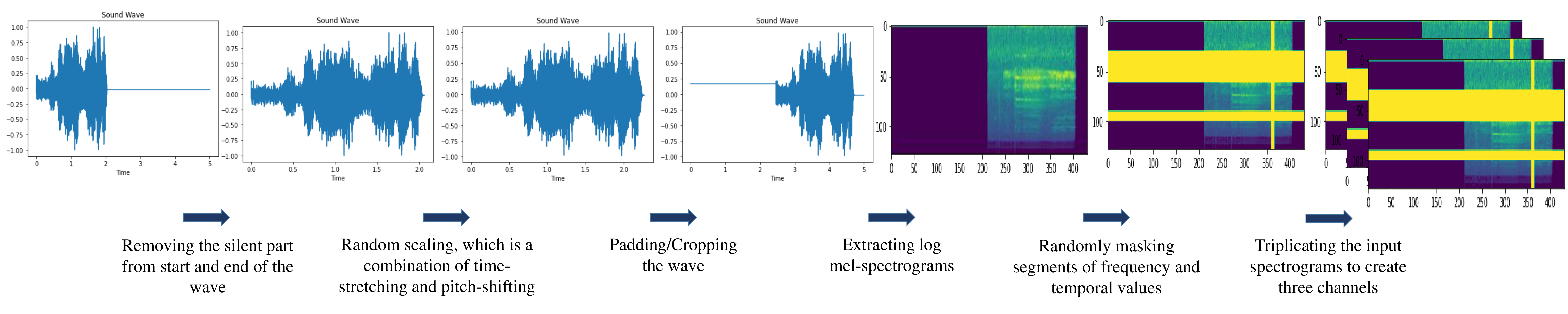}
  \caption{Data augmentation pipeline: The first three steps of removing the silent parts from the start and end of the signals, random scaling, and random padding/cropping is applied on the wave signals. Then we extract the log-scaled mel-spectrograms and randomly mask segments of the spectrograms across both time and frequency axes. At the end, we triplicate the masked spectrogram before inputting to the pre-trained model on the ImageNet.}
  \label{fig:dataAugment}
\end{figure*}

We design several augmentation techniques to be applied on the audio signals consecutively on both the wave and the spectrogram forms of the audio signals. Because of the randomness of these techniques, the same signal can be augmented differently at each training iteration. Figure \ref{fig:dataAugment} shows the order in which the data augmentation processes are applied on the samples. The techniques that we use on the wave form of the audio signals are:
\begin{enumerate}
    \item Removing Silent Beginning and End: We remove the silent parts from the start and the end of the signals. Some of the environmental sounds have a silent part in at least one end, which does not carry useful information. Removing the silent part allows more variety in the augmented signal, since the non-silent signal can be moved across the time axis more freely.
    
    \item Random Scaling: This is the same technique used in \cite{tokozume2018learning}, \cite{guzhov2020esresnet}. A random scaling factor is sampled uniformly from $[1.25^{-1}, 1.25]$, and is used for linear interpolation of the wave. This is equivalent to the combined pitch-shifting and time-stretching. We use this method because it only uses the waves and therefore is computationally much cheaper and faster than the time-stretching or pitch-shifting, where two short-time Fourier transforms are needed for each one of them.
    
    \item Random Padding / Random Crop: Removing the silent part and scaling the signal, changes its length. To have signals with uniform lengths, we use the random padding or cropping depending on the length of the modified signal.
\end{enumerate}
After applying these augmentations on the wave form of the audio, we extract the log-scaled mel-spectrograms from the waves. The log-scaled mel-spectrograms are two dimensional features which imitate the human hearing perception and are widely used in audio analysis. We apply another set of augmentations on these spectrograms including:
\begin{enumerate}
    \item Frequency Masking: We randomly select $f$ segments of the spectrograms in the frequency axis where width of each segment comes uniformly from $[0, F]$. The frequency values in these segments are set to $0$.
    \item Time Masking: We randomly select $t$ segments of the spectrograms in the time axis where width of each segment is drawn uniformly from $[0, T]$. The frequency values in these segments are set to $0$.
\end{enumerate}
By masking segments of the spectrograms along the time and frequency axes, we are forcing the model to focus more on the temporal-frequency patterns in order to classify them, rather than a specific frequency or temporal value. This is similar to how dropout acts in making the model to not rely on any specific neuron \cite{srivastava2014dropout}. 

\subsection{Baseline Model}
\label{baseline}
We use the network model $f(.)$ to create the representations for the augmented audio samples. We choose ResNet-50 \cite{he2016deep} as the representation model in all of our experiments. The output of the final average-pooling layer with 2048 dimensions in the ResNet model is normalized and we refer to it as the representation vectors for the audio samples. The classifier model is composed of a fully connected layer with $c$ hidden units, where $c$ specifies the number of the classes in the dataset. This classifier model uses the representation vectors as input, and outputs the probability of the samples belonging to each one of the $c$ classes. One-hot encoded vectors and cross-entropy loss are used to train the representation model and the classifier in an end-to-end manner. The loss is calculated as:
\begin{equation}
\mathcal{L}_{i}^{cross-entropy}  = - \sum_{i} y_{i}^{1-hot}  log(y_{i}^{logits})
\end{equation}
, where $y_{i}^{1-hot}$ is the 1-hot encoded ground-truth label for the sample $i$ and $y_{i}^{logits}$ is the output of the model for that sample.

\subsection{Training Audio Classifiers with Contrastive and Hybrid Loss}
\label{hybrid}
Contrastive loss is based on differentiation between positive and negative sample pairs. We want to maximize the similarity between the samples in the positive pairs and minimize the similarity between the samples in the negative pairs. We build the positive and negative pairs based on the labels of the samples, where samples with the same label in the mini-batch create the positives. Since all of the samples with the same label are grouped together, all of the samples in the mini-batch which are not in the same group with a given sample, are considered to be negatives in regard to that sample. 

Our supervised contrastive learning model consists of two networks: the representation network, and the projection network. The representation network $f(.)$ maps the sample $x_{i}$ to the representation vector $h_{i}$, where $h_{i} = f(x_{i}) \in \mathbb{R}^{D_{f}}$. Applying the contrastive loss on the representation vectors $h_{i}$, because of the high dimensionality of the representation space, does not separate the samples of the different classes from each other. Reducing the dimension of the representation space causes loss of information by shrinking the size of the representation vectors. To prevent shrinking the representation vectors and also to separate the samples of different classes from each other, we use an additional network, which we call the projection network. The projection network $g(.)$ performs dimension-reduction by linearly mapping the representation vectors to the projection space with lower dimension, $z(h_{i}) = g(h_{i}) \in \mathbb{R}^{D_{g}}$, where $D_{g} << D_{f}$. Our experiments show that applying the contrastive loss on the projection space, improves the performance of the classifier significantly.

 We run our experiments by using ResNet-50 as the representation network. Our projection model, which is a linear fully connected layer, uses the normalized representation vectors to output the projection vectors. The supervised contrastive loss \cite{khosla2020supervised}, is applied on the normalized projection vectors of the audio signals. Given labels $Y = \{y_{1}, y_{2}, ..., y_{N} \}$ corresponding to samples $X = \{x_{1}, x_{2}, ..., x_{N} \}$, the loss is calculated as:
\begin{equation}
    \mathcal{L}^{sup-cont}_{i} = - \frac{1}{N_{y_{i}}}log \frac{\sum_{j=1}^{N}\mathds{1}_{[y_{i}=y_{j}]}exp(z_{i}. z_{j}/\tau)}{\sum_{k=1}^{N} \mathds{1}_{[k\neq i]} exp(z_{i}. z_{k}/\tau)}
\end{equation}
where $N_{y_{i}}$ is the number of samples in the mini-batch with the same label as $y_{i}$, $\mathds{1}_{[ y_{i}=y_{j}]} \in \{0, 1\}$ evaluates to $1$ iff $y_{i}=y_{j}$, and $u.v$ measures the cosine similarity between the vectors $u$ and $v$ which is calculated as $u.v = u^{T}v/||u|| ||v||$. The $N$ refers to the batch size and $\tau$ is the temperature, which controls the effect of the hard negatives in the training process \cite{wang2020understanding}.

After training the representation and the projection networks with the contrastive loss, we freeze $f$ and train a classifier on the representation vectors, using the cross-entropy loss. In this way, the contrastive and the cross-entropy loss functions are being used to train the model in two separate stages, where the loss signal from the cross-entropy loss does not modify the parameters in the representation model. We believe these two losses can be applied as complementary to each other in one single training stage. In our hybrid approach, we propose to use the supervised contrastive loss to disentangle the representations of the samples of a class from the representations of the samples of the other classss, and to use the cross-entropy loss in mapping the logits of the samples of each class to the one-hot encoded vector. We believe that by combining these two loss signals, we can provide a stronger training signal from the representation model, which can lead to a better classification accuracy. Assuming that the cross-entropy loss identifies the similarities between the samples with the same label by mapping them to the same 1-hot encoded vector, and the contrastive loss tries to minimize the intra-class to inter-class distance ratio, combining these two loss signals provides the model with a much more informative signal for the classification task. In this hybrid method, we calculate the loss as:
\begin{equation}
\label{hybrid}
    \mathcal{L}_{i} = \alpha \mathcal{L}_{i}^{sup-cont}  + (1 - \alpha) \mathcal{L}_{i}^{cross-entropy}
\end{equation}
where $\alpha$ is a hyper-parameter. We show that using this combined loss, results in a faster and better convergence. To run the experiments using this loss function, we add two parallel branches to the output of the embedding network. As shown in Figure \ref{fig:overview}c, the logits from the last layer of the ResNet-50 are normalized to create the representation vectors. We apply two models in parallel to these vectors. One is the classifier which is a linear fully connected layer which has as many hidden units as the number of the classes and is used to calculate the cross-entropy loss. The other is a projection model with the same architecture as the classifier. The projection layer downsamples the representation vector and normalizes the output in order to calculate the supervised contrastive loss in a more efficient way within a lower dimension space.

\begin{table*}[h]
\caption{Accuracy of the baseline model on ESC-50 based on the different maximum widths and number of the masked segments
}
\label{tab:table_masks}
\centering

\begin{tabular}{c|cccc|cccc|cccc|}
\cline{2-13}
                            & \multicolumn{4}{c|}{F = 16, T = 32}                                                          & \multicolumn{4}{c|}{F = 32, T = 32}                                                                  & \multicolumn{4}{c|}{F = 32, T = 64}                                                         \\
                            & t = 0                      & t = 1                      & t = 2                      & t = 3 & t = 0                      & t = 1                              & t = 2                      & t = 3 & t = 0                      & t = 1                      & t = 2                     & t = 3 \\ \hline \hline
\multicolumn{1}{|c|}{f = 0} & \multicolumn{1}{c|}{90.2}  & \multicolumn{1}{c|}{90.35} & \multicolumn{1}{c|}{90.4}  & 90.45 & \multicolumn{1}{c|}{90.2}  & \multicolumn{1}{c|}{90.35}         & \multicolumn{1}{c|}{90.4}  & 90.45 & \multicolumn{1}{c|}{90.2}  & \multicolumn{1}{c|}{90.75} & \multicolumn{1}{c|}{90.6} & 90.35 \\ \hline
\multicolumn{1}{|c|}{f = 1} & \multicolumn{1}{c|}{91.2}  & \multicolumn{1}{c|}{91.6}  & \multicolumn{1}{c|}{91.35} & 91.1  & \multicolumn{1}{c|}{91.45} & \multicolumn{1}{c|}{91.7}          & \multicolumn{1}{c|}{91.5}  & 91.05 & \multicolumn{1}{c|}{91.6}  & \multicolumn{1}{c|}{91.85} & \multicolumn{1}{c|}{91.5} & 91.25 \\ \hline
\multicolumn{1}{|c|}{f = 2} & \multicolumn{1}{c|}{91.05} & \multicolumn{1}{c|}{91.35} & \multicolumn{1}{c|}{91.55} & 91.4  & \multicolumn{1}{c|}{91.6}  & \multicolumn{1}{c|}{\textbf{92.1}} & \multicolumn{1}{c|}{91.8}  & 91.4  & \multicolumn{1}{c|}{91.85} & \multicolumn{1}{c|}{91.65} & \multicolumn{1}{c|}{91.3} & 91.6  \\ \hline
\multicolumn{1}{|c|}{f = 3} & \multicolumn{1}{c|}{90.8}  & \multicolumn{1}{c|}{91.2}  & \multicolumn{1}{c|}{91.3}  & 91.55 & \multicolumn{1}{c|}{91.65} & \multicolumn{1}{c|}{91.7}          & \multicolumn{1}{c|}{91.45} & 90.95 & \multicolumn{1}{c|}{91.5}  & \multicolumn{1}{c|}{91.4}  & \multicolumn{1}{c|}{91.8} & 91.05 \\ \hline \hline
\end{tabular}
\end{table*}

\begin{figure*}[h]
  \centering
  
  \includegraphics[width=\textwidth]{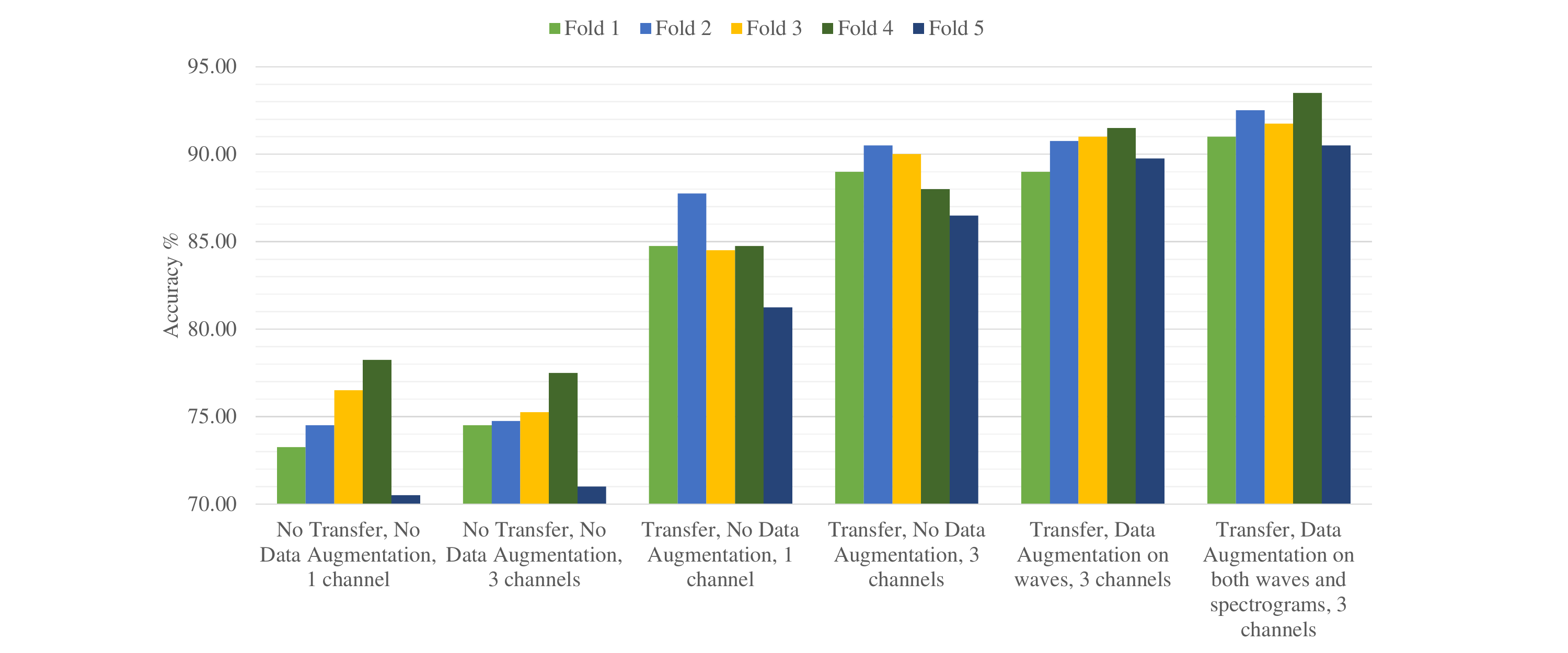}
  \caption{Results on application of transfer learning, data augmentation and number of input channels, by using the baseline model on ESC-50. Transfer learning has significant effect on improving the results. Increasing the number of the input channels from 1 to 3, improves the accuracies only when transfer learning is used. }
  \label{fig:chart_effects}
\end{figure*}

\section{Experiments and Results}
\label{experiments}
In this section, we describe the settings for our experiments, evaluate our proposed methods, and compare the results with the other methods in ESC.

\subsection{Datasets}
We train and evaluate our models using three publicly available datasets, which are commonly used in environmental sound classification studies. These datasets are ESC-50, ESC-10 \cite{piczak2015esc}, and UrbanSound8K \cite{salamon2014dataset}. 

The ESC-50 dataset includes 2000 audio samples (each is 5 seconds long), which are divided equally into 50 classes in 5 folds. Each fold is balanced regarding the number of the samples from different classes. This means that there are 8 samples for each class inside every fold. The classes can be divided into five major categories: animals, natural soundscapes and water sounds, non-speech human sounds, interior/domestic sounds, and exterior/urban sounds. The sampling rate of each audio signal is 44.1 kHz.

\begin{table}[]
\caption{ Results of different alpha values on the hybrid loss. The best results are achieved when the cross-entropy and contrastive loss contribute equally to calculating the loss signal.
}
\centering
\begin{tabular}{  P{1.5cm} | P{1.5cm}| P{1.5cm} |P{1.5cm} | } \cline{2-4}  \rule{0pt}{0.25cm}
                           & ESC-50 (\%)       & ESC-10 (\%)        & US8K (\%)           \\ \hline \hline 
\multicolumn{1}{|c|}{$\alpha = 0.01$} & 91.85         & 99.25         & 85.03          \\ \hline
\multicolumn{1}{|c|}{$\alpha = 0.25$} & 92.25         & 99.5 & 85.46          \\ \hline
\multicolumn{1}{|c|}{$\alpha = 0.5$}  & \textbf{93.2} & \textbf{99.75} & \textbf{85.96} \\ \hline
\multicolumn{1}{|c|}{$\alpha = 0.75$} & 92.8          & 99.25         & 84.94          \\ \hline
\multicolumn{1}{|c|}{$\alpha = 0.99$} & 92.7          & 99.25         & 84.72          \\ \hline \hline
\end{tabular}
\label{tab:alpha_vals}
\end{table}

The ESC-10 is a subset of ESC-50 with 400 samples in 10 classes. Like ESC-50, the ESC-10 is also divided into five folds. The classes in ESC-10 are dog bark, rain, sea waves, baby cry, clock tick, person sneeze, helicopter, chainsaw, rooster, and fire crackling. The length of each one of the sound samples is 5 seconds.

The UrbanSound8K or in short US8K, has 8732 samples from 10 different classes. The length of the audio clips are different and the maximum length is 4 seconds. The classes are air conditioner, car horn, children playing, dog bark, drilling, engine idling, gun shot, jackhammer, siren, and street music. The sound segments in US8K, have different sampling rates and can be mono or stereo. For the stereo audio segments, we calculate an average of the channels before augmenting the samples.

We use the official split of these three datasets in our experiments. 

\begin{table*}[h]
\caption{ Results of Our Three Methods. The average values shows the average validation accuracies of our models over 5 times training.
}
\centering
\begin{tabular}{cc|c|c|c|}
\cline{3-5}
                                                                    \rule{0pt}{0.25cm}          &         & ESC-50 (\%)  & ESC-10 (\%)  & US8K (\%)    \\ \hline \hline
\multicolumn{1}{|c|}{\multirow{2}{*}{ResNet-50, Cross-Entropy Loss}}          & Average & 91.85 $\pm$ 0.64   & 99.0 $\pm$ 0.47    & 85.28 $\pm$ 0.69 \\ \cline{2-5} 
\multicolumn{1}{|c|}{}                                                        & Best    & 92.65        & 99.5         & 86.16        \\ \hline
\multicolumn{1}{|c|}{\multirow{2}{*}{ResNet-50, Supervised Contrastive Loss}} & Average & 91.98 $\pm$ 0.68 & 99.25 $\pm$ 0.59 & 84.69 $\pm$ 0.98 \\ \cline{2-5} 
\multicolumn{1}{|c|}{}                                                        & Best    & 92.85        & 99.75        & 86.09        \\ \hline
\multicolumn{1}{|c|}{\multirow{2}{*}{ResNet-50, Hybrid Loss}}                 & Average & 92.86 $\pm$ 0.46 & 99.6 $\pm$ 0.14  & 85.78 $\pm$ 0.51 \\ \cline{2-5} 
\multicolumn{1}{|c|}{}                                                        & Best    & 93.4         & 99.75        & 86.49        \\ \hline \hline
\end{tabular}
\label{tab:table_threeModels}
\end{table*}

\begin{table*}[]
\caption{ Comparing the results of our proposed training method with the hybrid loss and other state-of-the-art methods. In order to have a fair comparison, we have separated the results achieved by the single models from the ensemble ones.
}
\centering
\begin{tabular}{|c|c|c|c|c|c|}
\hline \rule{0pt}{0.25cm}
Model               & Architecture                                   & Features                   & ESC-50 (\%)   & ESC-10 (\%)    & US8K (\%)      \\ \hline \hline
Human \cite{piczak2015esc}               & -                                               & -                          & 81.3          & 95.7           & -              \\ \hline
Piczak-CNN \cite{piczak2015environmental}         & 2 conv + 2 FC layers                           & Mel-Spectrogram            & 64.5          & 90.2           & 73.7           \\ \hline
EnvNet-v2 \cite{tokozume2018learning}          & 10 conv + 3 FC + 5 max-pooling layers          & Raw Wave                   & 84.7          & 91.3           & 78.3           \\ \hline
ESResNet \cite{guzhov2020esresnet}           & ResNet-50                  & Log Spectorgram     & 90.8          & 96.75          & 84.9           \\ \hline
ESResNet-Attention \cite{guzhov2020esresnet}  & ResNet-50 + Attention      & Log Spectrogram     & 91.5          & 97.0           & 85.42          \\ \hline
ResNet \cite{palanisamy2020rethinking}             & ResNet-50                  & Log Mel-Spectrogram & 90.93         & -              & 85.03          \\ \hline

DenseNet \cite{palanisamy2020rethinking}           & DenseNet-201              & Log Mel-Spectrogram & 91.52         & -              & 85.31          \\ \hline
SoundCLR & ResNet-50                  & Log Mel-Spectrogram & \textbf{93.4} & \textbf{99.75} & \textbf{86.49}          \\ \hline \hline
ResNet-Ensemble \cite{palanisamy2020rethinking}    & Ensemble of 5 ResNet-50    & Log Mel-Spectrogram & 92.64         & -              & 87.35          \\ \hline
DenseNet-Ensemble \cite{palanisamy2020rethinking}  & Ensemble of 5 DenseNet-201 & Log Mel-Spectrogram & 92.89         & -              & 87.42 \\ \hline
SoundCLR-Ensemble  & Ensemble of 5 ResNet-50 & Log Mel-Spectrogram & \textbf{93.6}         & \textbf{99.75}              & \textbf{88.01} \\
\hline \hline

\end{tabular}
\label{tab:table_otherMethods}
\end{table*}

\subsection{Preprocessing and Data Augmentation}
We normalize all of the wave signals into values in the range of [-1, 1]. Then in training, we make the signals go through the data augmentation pipeline described in section \ref{augmentation}. For ESC-10 and ESC-50, we perform the random padding and random crop with the output size of 22500 to include 5 seconds of data. For US8K, this output size is 176400 to have signals with length of 4 seconds which is the maximum length for the audio clips in this dataset.

To calculate the mel spectrograms, we use hamming window with a size of 1024 with 50\% overlap and 128 mel-bands. We get the logarithm of the mel-spectrograms to make our features more compatible with the humans hearing sense. 

We are augmenting the data in two stages: wave, and spectrogram. We explained the techniques used in each stage in \ref{augmentation}. To show the contribution of the different modules in our augmentations pipeline, we run experiments on ESC-50 using our baseline model and show the results in Figure \ref{fig:chart_effects}. Our first experiment is run by randomly initializing ResNet-50, without using any of the data augmentation techniques, and feeding the mel-spectrograms in one channel to the model which achieves an average validation accuracy of 74.6 \% across all 5 folds of ESC-50. Triplicating number of channels at this point (not using transfer learning) does not improve the results. By not using the transfer learning, increasing the number of channels just increases the size of the input without providing any extra information to the model and the randomly initialized model cannot utilize this extra redundant information. Initializing the weights of the model with the parameters learnt by training on the ImageNet, increases the average accuracy from 74.6\% to 84.6\%. This significant rise in the accuracy is due to the correlation among the nearby spectrotemporal values in the spectrograms which is similar to the correlation among the nearby pixels in an image. Since the parameters of the ResNet-50 model that we use as our feature extractor, is trained on images with three channels, to utilize all of the parameters of the model, we present our input in three channels. By simply copying the log-scaled mel-spectrogram values to increase the number of the channels to three. As shown in Figure \ref{fig:chart_effects}, this simple technique increases the average accuracy value by more than 4\%. Each set of the augmentations on the wave signals and mel-spectrograms, as outlined in section \ref{augmentation} and shown in Figure \ref{fig:chart_effects}, further improve the average accuracy by 1.6\% and 1.45\%, respectively.

To identify the best parameters for the number of the masked segments ($f$ and $t$) and their maximum-width ($F$ and $T$), we run experiments using values of $\{16, 32\}$ for $F$ and $\{32, 64\}$ for $T$, where $F$ and $T$ are the maximum width of each frequency and temporal masked segments. To find the best number for the masked segments across the frequency and time axes, we perform a grid-search for $f, t \in \{0, 1, 2, 3 \}$, where $f$ is number of the masked segments in frequency values and $t$ is the number of masked temporal segments. Table \ref{tab:table_masks} shows the accuracy values from running the baseline model on ESC-50. It turns out that the best accuracy for the baseline model happens when $F = 32, T = 32$ and $f = 2, t = 1$. Using these settings for masking, improves the accuracy of the baseline model on ESC-50 from 90.2\% to 92.1\%. We believe that this improvement is caused by forcing the model not to be dependent on any specific spectrotemporal values. Our results show that the frequency masking is comparatively more effective than the temporal masking. This might be in part due to the higher importance of the information in the frequency bands compared to most of the short temporal segments, which forces the model to identify more global patterns in classifying the audio events. Increasing the number of the masked segments beyond two frequency and one temporal segments, eliminates some of the essential information and reduces the overall accuracy of the model.

\begin{figure}
  \centering
  
  \includegraphics[width=0.45\textwidth]{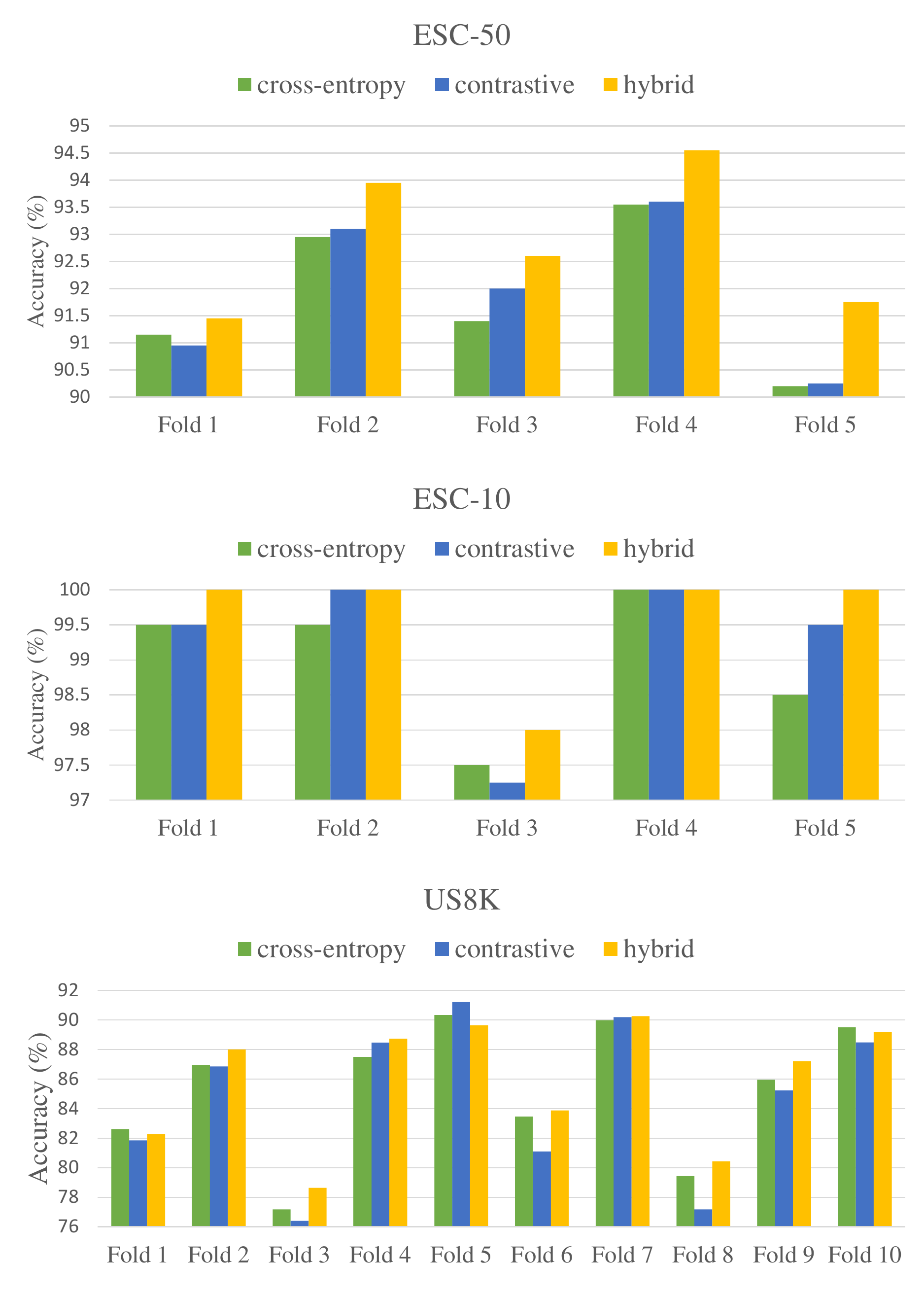}
  \caption{The validation accuracies over different folds of the datasets averaged over 5 times training. In most of the folds, the models trained with the hybrid loss reach higher accuracies that the ones trained with the cross-entropy and contrastive loss functions. }
  \label{fig:average_folds}
\end{figure}

\begin{figure*}[h]
  \centering
  \includegraphics[width=\textwidth]{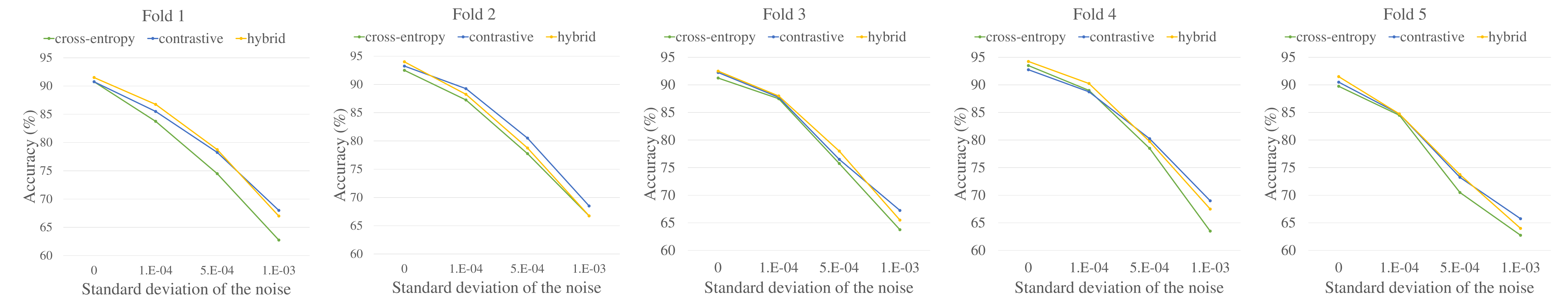}
  \caption{Effect of white noise on the accuracy values of the classifiers trained with cross-entropy, supervised contrastive, and hybrid loss on ESC-50. The models trained with the supervised contrastive loss show more resilience towards the noise and the accuracies of the models trained with this loss decreases slower than the models trained with the other two losses.}
  \label{fig:noise}
\end{figure*}

\subsection{Training Settings}
We use ResNet-50 as the representation model in our experiments. The outputs of the final pooling layer with 2048 dimension is normalized to generate the representation vector for each sample. In the baseline model we add a linear fully connected layer with $h_c$ hidden units where $h_c$ is the number of the classes in the dataset. The $h_c = 10$ for ESC-10 and US8K and it is 50 for ESC-50. We train the baseline model using the cross-entropy loss.

In training with supervised contrastive loss, we use a linear projection model with one fully connected layer with linear activation. Based on our experiments using a linear layer with 64 hidden-units, achieves the best accuracy as it makes the representation space small-enough to be able to efficiently separate the samples from each other, and yet large enough to preserve the most important features. For the supervised contrastive loss, we use the implementation from \cite{khosla2020supervised}. We use the samples in the mini-batches to draw the positives and negatives for each sample in the batch. It has been shown both in self-supervised and supervised contrastive loss \cite{chen2020simple}, \cite{khosla2020supervised}, that having larger number of negatives can help the training process. For this reason, we use batches of 128 samples to be able to include more samples in the positive and negative groups and create a more informative contrast. In the case of ESC-50, for each sample, we have more than 2 samples from the same class (positive samples), and more than 125 samples from other classes (negative samples) in average.  

We used our baseline model to perform a grid-search to identify the optimal hyper-parameters for the training stage. We use Adam optimizer with a base learning rate of $5e-4$ for ESC-50 and ESC-10, and $1e-4$ for US8K. We also applied an exponential learning rate decay with a factor of 0.98 and 10 warm-up epochs. To ensure a fair comparison among our models, we use the same hyperparameters in training the baseline, models with supervised contrastive loss, and the hybrid loss.

\subsection{Results of Our Three Models}
\label{results}
We run experiments to identify the optimal value for $\alpha$ in Formula \ref{hybrid}, from the possible values of $\{0.01, 0.25, 0.5, 0.75, 0.99\}$. When $\alpha=0.01$, most of the hybrid loss value comes from the cross-entropy loss, and when $\alpha = 0.99$, the supervised contrastive loss is the dominant factor. As shown in Table \ref{tab:alpha_vals}, the highest accuracy is achieved when $\alpha = 0.5$ for all three datasets. This shows that by equal contribution from the cross-entropy and contrastive loss functions, we can achieve higher accuracy and the model can achieve higher accuracy when these two loss functions act as complementary to each other.

We also compare the average of five runs and the best results of our three models trained with cross-entropy, supervised contrastive, and hybrid loss with each other in Table \ref{tab:table_threeModels}. On average, ResNet-50 model trained with our proposed hybrid loss achieves validation accuracies of 92.86\%, 99.6\%, and 85.78\% on ESC-50, ESC-10, and US8K. This shows that not only by combining the two stages of training in the supervised contrastive learning we can make the training faster, but also the hybrid loss actually increases the performance of the model by providing a stronger loss signal. In Figure \ref{fig:average_folds}, we show the average performance of our models on each fold in the experimented datasets. This figure shows that the model trained with the hybrid loss either outperforms or have the same accuracy across all of the folds in ESC-50 and ESC-10, and most of the folds in US8K.

In Table \ref{tab:table_otherMethods}, we compare our hybrid approach, which is based on training a single ResNet-50 model with the hybrid loss, with other methods in ESC. Since in \cite{palanisamy2020rethinking}, an ensemble of ResNet and DenseNet models are used in ESC, to have a fair comparison, we calculate the performance of an ensemble of five independently trained ResNet-50 models based on our proposed hybrid loss. Here, we use the average of the softmax output of these five models to get the accuracy value for the ensemble model.

As shown in Table \ref{tab:table_otherMethods}, a single ResNet-50 trained with the hybrid loss, increases the accuracies on ESC-50, ESC-10, and US8K by 1.88\%, 2.75\% and 1.18\% respectively compared to the previous state-of-the-art results from DenseNet in \cite{palanisamy2020rethinking}. The improvement is more significant, if we compare our results with the ResNet-50 model trained in \cite{palanisamy2020rethinking}. We believe that our data augmentation techniques and use of the contrastive loss has a large role in achieving these state-of-the-art results. Ensembling our five ResNet-50 models increases the accuracy over ESC-50 and US8K datasets to 93.6\% and 88.01\%.

\subsection{Robustness to Noise}
\label{subsec:noise_effect}
The environmental sound classifiers need to be robust to the noise, since environmental sound recordings usually include several types of background noise. We use white noise signal with normal distribution to study the effect of the noise in our classifiers. These noise signals are created randomly with mean of 0 and different standard deviation values of $1e-4$, $5e-4$, and $1e-3$. We add the noise signals to the audios in the test set. Figure \ref{fig:noise}, shows the effect of the noise on the performance of our classifiers trained on ESC-50 dataset. This figure shows that the models trained with the supervised contrastive loss are more resistant against the noise and the accuracies of these model decreases slower than the other two models. We think this is due the higher distance between the samples in the representation space of the models trained with the contrastive loss. The contrastive loss is based on increasing the distance among the samples of the different classes and this increased margin contributes to the better performance on the noisy signals.

\section{Conclusion}
\label{conclusion}
In this paper, we introduced SoundCLR, a novel framework based on strong data augmentation and contrastive learning to classify environmental audio events. We introduced the supervised contrastive learning concept into ESC, and showed that our model can achieve state-of-the-art results in ESC by using the contrastive loss to disentangle the samples of the different classes from each other in the representation space, combined with the cross-entropy loss used to map the representation vectors to the output labels. In our proposed method, we demonstrated how the contrastive and cross-entropy loss functions can be used as complementary to each other in audio classification to provide a stronger error signal in training the audio classifiers. Furthermore, we showed how simple data augmentation techniques such as masking and increasing the number of the channels in the input mel-spectrograms can significantly increase the classification accuracy of the model.

\section{Acknowledgement}
Research reported in this work was supported in part by NSF under grant and 1940099 and 1905775. The views, perspective, and content do not necessarily represent the official views of the NSF.


\end{document}